\newcommand{\beqa}{\begin{eqnarray*}}
\newcommand{\eeqa}{\end{eqnarray*}}
\newcommand{\beqan}{\begin{eqnarray}}
\newcommand{\eeqan}{\end{eqnarray}}
\newcommand{\beq}{\begin{equation}}
\newcommand{\eeq}{\end{equation}}
\newcommand{\diff}{{\rm d}}
\newcommand{\drr}{\frac{\partial}{\partial r}}
\newcommand{\dtt}{\frac{\diff}{\diff t}}
\newcommand{\dr}[1]{\frac{\partial  #1}{\partial r}}

\newcommand{\lp}{ \left(}
\newcommand{\rp}{ \right)}
\newcommand{\lc}{ \left[}
\newcommand{\rc}{ \right]}
\newcommand{\cf}{{\it cf.}}

\documentclass{iaus}
\usepackage{graphicx}

\title[Deep inside low-mass stars] 
{Deep inside low-mass stars}

\author[Charbonnel \& Talon]   
{Corinne Charbonnel $^{1,2}$, Suzanne Talon$^3$}

\affiliation{$^1$ Geneva Observatory, University of Geneva, ch. des Maillettes 51, 1290 Versoix, Switzerland \break email: Corinne.Charbonnel@obs.unige.ch \\[\affilskip]
$^2$ Laboratoire d'Astrophysique de Toulouse-Tarbes, Universit\'e de Toulouse, CNRS UMR 5572, 14 Av. E.Belin, 31400 Toulouse, France \\[\affilskip]
$^3$ RQCHP, D\'epartement de physique, Universit\'e de Montr\'eal, C.P. 6128, succ. centre-ville, Montr\'eal, Qu\'ebec, H3C 3J7  \break email: talonsuz@rqchp.qc.ca}

\pubyear{2008}
\volume{252}  
\pagerange{1--12}
\date{?? and in revised form ??}
\jname{Proceedings ``The Art of Modelling stars in the 21st Century" - IAU Symposium}
\editors{L.Deng, K.L.Chan \& C.Chiosi, eds.}
\begin{document}

\maketitle

\begin{abstract}
Low-mass stars exhibit, at all stages of their evolution, the signatures 
of complex physical processes that require challenging modeling
beyond standard stellar theory.
In this review, we recall the most striking observational evidences 
that probe the interaction and interdependence of various transport processes 
of chemicals and angular momentum in these objects. We then focus on the impact
of atomic diffusion, large scale mixing due to rotation, and internal gravity
waves on stellar properties on the main sequence and slightly beyond.

\end{abstract}

\firstsection 
\section{Clues on transport processes in low-mass stars}
During the last couple of decades, it became obvious that ``the art of 
modeling stars in the 21st century" will actually strongly rely on the art 
of modeling transport processes in stars. 
Observational evidences now give precise clues on the various processes that 
transport angular momentum and chemical elements in the radiative regions 
of low-mass stars, at various phases of their evolution. 
Here are a few examples of
observations that require modeling beyond standard stellar theory: 
The abundance patterns of lithium in the Sun, in field and cluster main sequence 
and subgiant stars, as well as in Population II dwarfs; the rotation profile in the Sun 
inferred from helioseismology; the abundance patterns of lithium, beryllium, carbon 
and nitrogen on the red giant branch (see Charbonnel \& Zahn 2007a, b, and this volume); 
the intrinsic s-process elements observed in AGB stars; the abundance of helium 3
and heavier elements in planetary nebulae; the observed spin of white dwarfs.

One of the most striking signatures of transport processes in low-mass stars
is the so-called Li dip (see Fig.~\ref{VLiHyades}).
This drop-off in the Li content of main-sequence F-stars
in a range of $\sim 300~{\rm K}$ centered around $6700~{\rm K}$ was discovered in the
Hyades by Wallerstein et al. (1965); its existence was latter confirmed by
Boesgaard \& Tripicco (1986).
This feature appears in all open clusters older than $\sim 200~{\rm Myr}$, 
as well as in field stars (Balachandran 1995), an indication that it is a phenomenon 
occurring on the main sequence.

It was first suggested by Michaud (1986) that the Li dip could be due to element
separation below the stellar convective envelope.
He showed that Li is supported
by radiative acceleration at the bottom of the convective envelope in stars
with $T_{\rm eff} \geq 7000~{\rm K}$, while gravitational settling dominates and leads
to Li underabundances in stars with $T_{\rm eff}$ between $6800$ and $6400~{\rm K}$.
In cooler stars, at the age of the Hyades,
atomic diffusion did not have enough time to modify the Li abundance in the deep
surface convection zone.
These predictions were obtained from first principles, the only free parameter
being the mixing length parameter, which controls the effective temperature
of the Li dip. In addition, a mass loss rate of the
order of $10^{-15}M_{\odot}\,{\rm yr}^{-1}$ was required to reduce the
expected Li overabundance on the hot side of the dip.
More sophisticated models based solely on atomic diffusion were constructed 
by Richer \& Michaud (1993),
but this explanation suffers from two serious drawbacks: \\
~$\blacktriangleright$~ 
The expected concomitant  underabundances of heavier elements (C, N, O, Mg, Si) 
are not observed in cluster stars 
(Takeda et al. 1998; Varenne \& Monier 1999; Gebran et al. 2008). \\
~$\blacktriangleright$~ In this framework Li is not destroyed;
it rather settles out of the convective envelope and accumulates in a buffer zone below. 
As a consequence, Li should be dredged-up as a star enters the Hertzsprung gap.
This is not seen in the Li data, neither in the field nor in open cluster stars
(Pilachowski et al. 1988; Deliyannis et al. 1997). 

This suggests that another process is responsible for the existence of the Li dip.
Boesgaard (1987) noticed that the effective temperature of the dip is also associated 
with a sharp drop in rotation velocities as can be seen in Fig.~\ref{VLiHyades}.
Rotation was then suggested to play a dominant role in this mass range.

\section{Rotation-induced transport in low-mass stars}

In order to properly model rotation-induced mixing, one must follow 
the time evolution of the angular momentum distribution within a star, 
taking into account {\em all} relevant physical processes:
contraction/expansion caused by the stellar evolution, mass loss 
or accretion, tidal effects, as well as internal redistribution of angular momentum 
through meridional circulation, turbulence, magnetic torques, waves, etc.

As discussed by Zahn in these proceedings, the description of the internal physical 
processes related to stellar rotation has been greatly improved during the last 
two decades. 
Talon \& Charbonnel (1998; see also Charbonnel \& Talon 1999, Palacios
et al. 2003, Pasquini et al. 2004, and Decressin et al. in preparation) showed that 
the hot side of the Li dip (down to $T_{\rm eff}\sim 6500~{\rm K}$) is very well 
reproduced using Zahn's (1992) model of rotation-induced mixing and the same free 
parameters as those required to explain abundance anomalies in more massive stars
(see Zahn, and Meynet, this volume).
In this framework, transport of angular momentum is dominated by the Eddington-Sweet
meridional circulation and shear instabilities.
The blue side of the Li dip is then attributed to enhanced mixing (and thus, Li burning)
caused by the large angular momentum gradients created by surface braking.
However these rotating models fail to reproduce the Li rise on the cool side of the dip, 
i.e., for stars with effective temperature lower than $\sim 6600~{\rm K}$. Why is it so?

\begin{figure}
\centering
\includegraphics[width=4cm,angle=0]{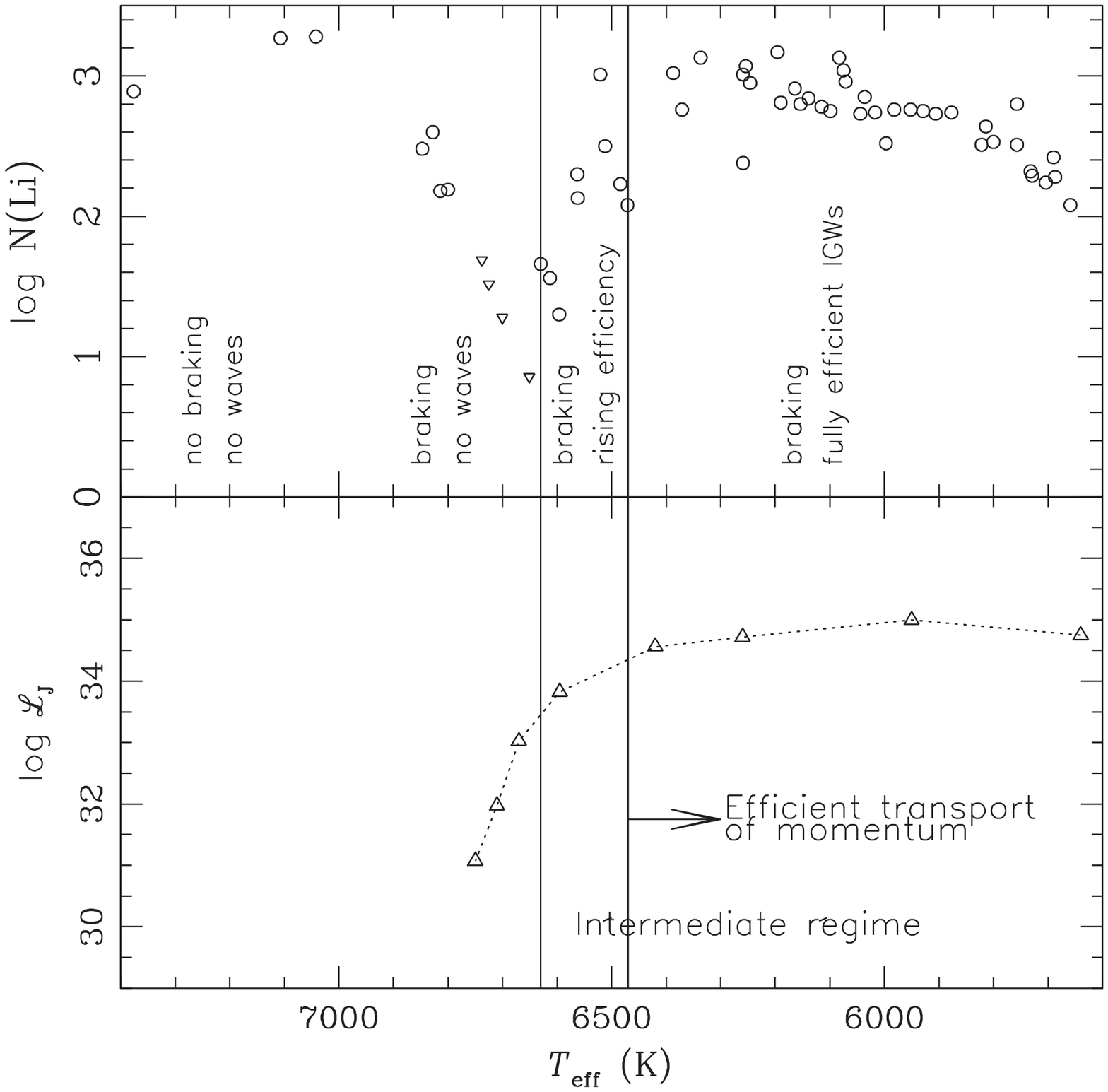}\qquad
\includegraphics[width=4cm,angle=0]{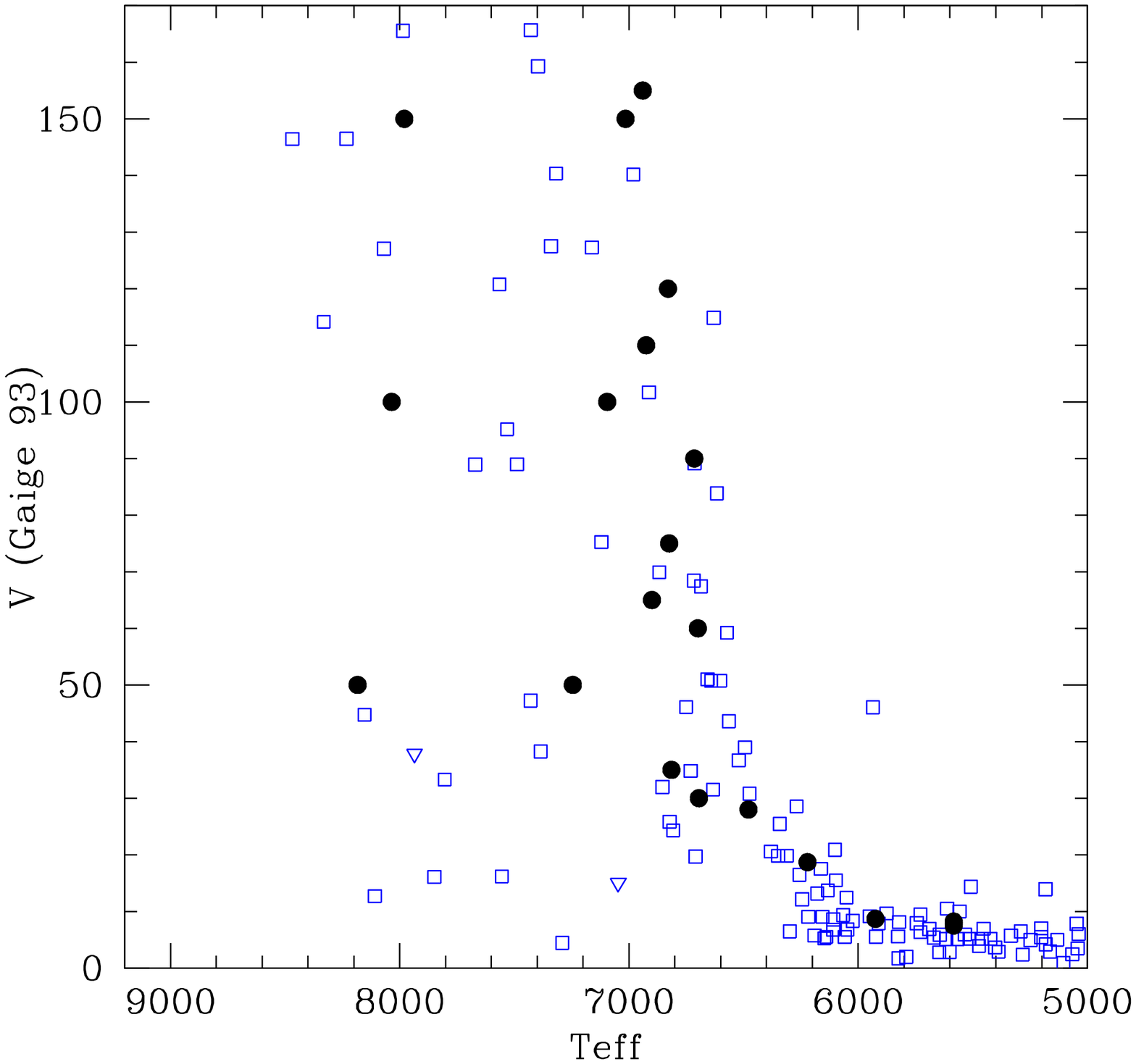}\qquad
\includegraphics[width=4cm,angle=0]{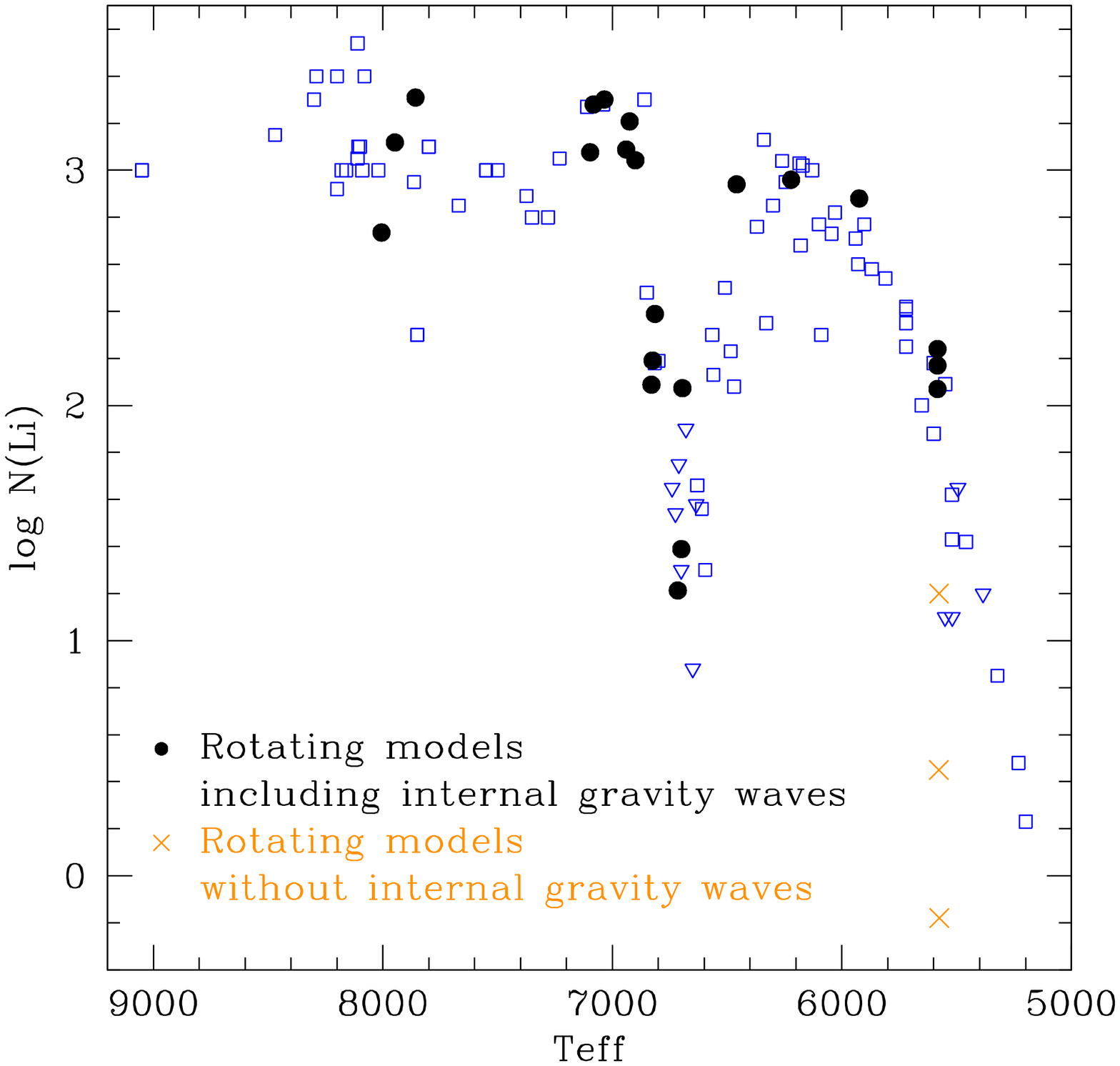}
  \caption{
{\bf(Left)} ({\em top}) Lithium abundance in the Hyades versus effective
  temperature. Also indicated are the approximate dependence of surface braking on
  effective temperature and the requirements for angular momentum
  transport so that rotational mixing can be expected to lead to the formation of
  the lithium dip (Talon \& Charbonnel~1998). ({\em bottom})
  Filtered angular momentum luminosity of waves below the SLO measuring the efficiency
  of wave induced angular momentum extraction.
  Adapted from Talon \& Charbonnel~(2003).
{\bf(Middle)}
Projected rotational velocity in the Hyades (Gaig\'e 1993; open squares and triangles).
{\bf(Right)} Li abundances versus effective temperature in the Hyades, Coma B and
Praesepe  (Burkhart \& Coupry 1998, 2000; open squares and triangles).
Comparison with our models including IGWs (black dots; Charbonnel et al. in preparation);
the crosses indicate the predictions for the 1~M$_{\odot}$ computed without IGW}
\label{VLiHyades}
\end{figure}

This effective temperature appears to be a ``pivotal" value 
in the rotation history of stars (see \,Fig.~\ref{VLiHyades}).
Indeed, the physical processes responsible for the evolution of the surface velocity
are different on each side of the Li dip, and the plot may be split into three 
temperature ranges associated with different dominant physical processes:\\
 ~$\blacktriangleright$~ Stars with $T_{\rm eff}$ higher than $\sim 6900~{\rm K}$ have
a very shallow convective envelope, which is not an efficient site for
magnetic generation via a dynamo process\footnote{Here, we discuss the presence
of an ``external" magnetic field, which could interact with mass loss.}. 
Thus, contrary to Sun-like stars, they are not slowed down by a magnetic torque. 
As a result, these stars soon reach a stationary regime where there is 
no net flux of angular momentum\footnote{In fact, there remains a small flux of 
angular momentum, just sufficient to counteract the effect of stellar 
contraction/expansion.}, in which meridional circulation and turbulence 
counterbalance each other. The associated weak mixing is just sufficient to counteract 
atomic diffusion. These rotating models account nicely for the observed constancy 
of Li and CNO in these stars, and they also explain the Li behaviour in subgiant 
stars (Palacios et al. 2003; Pasquini et al. 2004). 

\noindent In this region Am stars, which are known to be slow rotators, offer very 
good constraints to probe the processes that compete with atomic diffusion 
(e.g., Richer et al. 2000). Talon, Richard, \& Michaud (2006) showed
that the transport coefficients related to rotation-induced mixing 
lead to normal A stars for rotation velocities above $\sim 100~{\rm km \,s^{-1}}$, 
and permit Am anomalies below, with a mild correlation with rotation, provided 
a reduction of turbulent mixing by horizontal turbulence is taken into account. 
Fossati et al. (2008) have confirmed observationally this correlation with rotation:
in a large sample of stars in Praesepe, they find indeed 
a strong correlation between chemical pecularities of Am stars and their $v\sin i$. \\
 ~$\blacktriangleright$~ Between $\sim 6900$ and $6600~{\rm K}$, the convective envelope 
deepens and a weak magnetic torque, associated with the appearing dynamo, spins
down the outer layers of the star. In this case, the transport of angular momentum 
by meridional circulation and shear turbulence increases, leading to a larger 
destruction of Li, in agreement with the data. The rotating model thus perfectly 
fits the blue side of the Li dip. \\
 ~$\blacktriangleright$~ Stars on the cool side of the Li dip 
($T_{\rm eff} < 6600~{\rm K}$) have an even deeper convective envelope sustaining 
a very efficient dynamo, which produces a strong magnetic torque that spins down the
outer layers very efficiently. If we assume that all the angular momentum transport 
is assured by the wind-driven circulation in these 
stars\footnote{Let us mention
that in the case of large shears meridional circulation is far more efficient than shear
for angular momentum transport.}, we obtain too much Li depletion compared to the 
observations 
(see Fig.~\ref{evo_li}). 
On the basis of these results, Talon \& Charbonnel (1998) suggested that the Li dip 
corresponds to a transition region where another internal process starts to efficiently
transport angular momentum. 

This proposition is directly linked to another observation that fails to be reproduced 
by the pure hydrodynamic models, namely the flat solar rotation profile revealed by 
helioseismology (Brown et al. 1989; Kosovichev et al. 1997; Couvidat et al. 2003; 
Garcia et al. 2007; see also Christensen-Daalsgard, this volume).
At the solar age indeed, models relying only on turbulence
and meridional circulation for momentum transport predict large
angular velocity gradients that are not present in the Sun
(Pinsonneault et al. 1989; Chaboyer et al. 1995; Talon 1997; Matias \& Zahn 1998). 
This is an additional clue that another process participates to the transport of
angular momentum in solar-type stars.

Two mechanisms have been proposed to explain the near uniformity of the solar 
rotation profile.  The first rests on the possible existence of a magnetic field 
within the radiation zone (Charbonneau \& MacGregor~1993; Eggenberger et al.~ 2005).
The second invokes traveling internal gravity waves (hereafter IGWs) generated 
at the base of the convection envelope (Schatzman~1993; Zahn et al.~1997; 
Kumar \& Quataert~1997; Talon et al.~2002).
For either of these solutions to be convincing, they must be tested with numerical models
coupling these processes with rotational instabilities and should explain all the aspects
of the problem, including the lithium evolution with time.

In the case of the magnetic field, two different models have been suggested.
First, in a series of 2--D numerical calculations based on a static poloidal field 
fully contained in the Sun's radiative zone, it has been shown that, at low latitudes, 
radial differential rotation can be severely limited (Charbonneau \& MacGregor~1993;
MacGregor \& Charbonneau~1999). Shear is also reduced in these models, and, for the solar 
case, lithium burning could be reconciled with observations
(Barnes et al. 1999). The temperature dependence has not been examined
in the framework of that model, and no natural relation between the existence of a surface
convection zone and an internal poloidal field is expected. Furthermore, recent
numerical simulations by Brun \& Zahn~(2006) have cast doubt on the possibility that such 
a field would remain confined to the radiative zone for over $4\,{\rm Gyr}$
(see however the recent results by Wood \& McIntyre 2007). 
Second, Eggenberger et al.~(2005) studied the impact of the Tayler-Spruit instability in
solar models. The impact on light element abundances has not been quantified 
and once more, no correlation is expected with surface temperature.

The Li data suggest that the efficiency of the additional process is linked to the 
growth of the convective envelope in stars with effective temperatures around
$T_{\rm eff} \simeq 6600~{\rm K}$. As we shall see below, this is a characteristic of IGWs.
 
\section{IGWs generation and momentum extraction in low-mass stars}
In the Earth's atmosphere, wave-induced momentum transport is a key process in the 
understanding of several phenomena,
the best known being the quasi-biennial oscillation of the stratosphere.
In astrophysics, IGWs have initially been invoked as a source of mixing for 
chemicals (Press 1981; Garcia Lopez \& Spruit 1991; Schatzman 1993; Montalban 1994; 
Montalban \& Schatzman 1996, 2000; Young et al. 2003). Ando (1986) studied the 
transport of angular momentum associated with standing gravity waves in Be stars. 
He was the first to clearly state, in the stellar context, that IGWs carry angular 
momentum from the region where they are excited to the region where they are 
dissipated. Traveling IGWs have since been invoked as an important source of 
angular momentum redistribution in single stars (Schatzman 1993; Kumar \& Quataert 
1997; Talon et al. 1997, 2002; Charbonnel \& Talon 2005). 

\subsection{IGWs generation and wave spectrum}

The existence of IGWs in stars is expected from numerical simulations of penetrative
convection both in 2-- and 3--D (Hurlburt et al. 1986, 1994; Andersen 1994; Nordlund 
et al. 1996; Kiraga et al. 2000; Dintrans et al. 2005; Rogers \& Glatzmaier 
2005a, b). Ultimately, one may wish to obtain realistic wave fluxes from such 
simulations; however, conditions for these simulations are still too far from realistic
to be used at this time (for more details on these aspects see Charbonnel \& Talon 2007 
and the discussion at the end of the present paper). From a theoretical viewpoint,
two different and complementary processes excite IGWs: convective overshooting
in a stable region (Garc\'{\i}a L\'opez \& Spruit 1991; Kiraga et al. 2003;
Rogers \& Glatzmaier 2004), and Reynolds stresses in the convection zone 
(Goldreich \& Kumar 1990; Balmforth 1992; Goldreich et al. 1994). Here, we shall use 
the second mechanism, which has been calibrated on solar p-modes and, thus, seems 
more reliable at this time. We are fully aware that, as of
now, this is the weakest point of wave modeling.
Work in underway to evaluate analytically the contribution of penetrative plumes
to excitation (Belkacem et al.~2008).

For the evolution of angular momentum deep inside the star, the relevant parameter is
the net angular momentum luminosity slightly below the convective envelope. To estimate
that luminosity, we first need to estimate the spectrum of excited waves. 
In the Goldreich et al. (1994) model that we use, driving is dominated by entropy 
fluctuations.  In our calculations
(Talon \& Charbonnel 2003, 2004, 2005, 2007, 2008; Charbonnel \& Talon 2005), 
we neglect wave generation by overshooting, although it 
is expected to be very efficient; our present estimate is thus a lower limit to the 
correct/total wave flux.
As far as numerical simulations are concerned, some authors agree as to the order of 
magnitude we obtain in our wave-flux (Kiraga et al.~2000; Dintrans et al.~2005) while
Rogers \& Glatzmaier (2005) state that we over-estimate this flux. The reason for this 
discrepancy has been discussed in Charbonnel \& Talon (2007), and we are confident that, 
as 3--D numerical simulations evolve to more realistic regimes, the simulated wave flux 
will rise with turbulence.

In Talon \& Charbonnel~(2005), we developed a formalism to incorporate the contribution
of IGWs to the transport of angular momentum and chemical elements in stellar models.
We showed that the development of a double-peaked shear layer (SLO, for Shear Layer
Oscillation), acts as a filter for waves and discussed how the asymmetry of this filter 
produces momentum extraction from the core when it is rotating faster than the surface. 
Using only this filtered flux, it is possible to follow the contribution of internal 
waves over long (evolutionary) time-scales.
Let us recall the main features of our formalism.
The energy flux per unit frequency ${\cal F}_E$ is 
\begin{eqnarray}
{\cal F}_E \lp \ell, \omega \rp &=& \frac{\omega^2}{4\pi} \int dr\; \frac{\rho^2}{r^2}
   \left[\left(\frac{\partial \xi_r}{\partial r}\right)^2 +
   \ell(\ell+1)\left(\frac{\partial \xi_h}{\partial r}\right)^2 \right]  \nonumber \\
 && \times  \exp\left[ -h_\omega^2 \ell(\ell+1)/2r^2\right] \frac{v_c^3 L^4 }{1
  + (\omega \tau_L)^{15/2}},
\label{gold}
\end{eqnarray}
where
$\xi_r$ and $[\ell(\ell+1)]^{1/2}\xi_h$ are the radial and horizontal
displacement wave-functions, which are normalized to unit energy flux just
below the convection zone, $v_c$ is the convective velocity,
$L=\alpha_{\rm MLT} H_P$ the radial
size of an energy bearing turbulent eddy, $\tau_L \approx L/v_c$ the
characteristic convective time, and $h_\omega$ is the
radial size of the largest eddy at depth $r$ with characteristic frequency of
$\omega$ or greater ($h_\omega = L \min\{1, (2\omega\tau_L)^{-3/2}\}$).
The radial wave number $k_r$ is related to the horizontal wave number $k_h$ by
\beq
k_r^2 = \lp \frac{N^2}{\sigma^2} -1 \rp k_h^2 =
\lp \frac{N^2}{\sigma^2} -1 \rp \frac{\ell \lp \ell +1 \rp}{r^2} \label{kradial}
\eeq
where $N^2$ is the Brunt-V\"ais\"al\"a frequency. In the convection zone, the mode is
evanescent and the penetration depth varies as $\sqrt{\ell \lp \ell +1 \rp}$.
The momentum flux per unit frequency ${\cal F}_J$ is then related to the energy flux by
\begin{equation}
{{\cal F}_J}\lp m, \ell, \omega \rp = \frac{2m}{\omega} {\cal F}_E\lp \ell, \omega \rp
\end{equation}
(Goldreich \& Nicholson~1989; Zahn, Talon \& Matias~1997).
We integrate this quantity horizontally to get an angular momentum luminosity
\beq
{\cal L}_J = 4 \pi r^2 {\cal F}_J
\eeq
which, in the absence of dissipation, is conserved (Bretherton~1969; Zahn et al.~1997).
Each wave then travels inward and is damped by thermal diffusivity and by viscosity.
The local momentum luminosity of waves is given by
\beq
{\cal L}_J(r) = \sum_{\sigma, \ell, m} {{\cal L}_J}_{\ell, m} \lp r_{\rm cz}\rp
\exp \lc -\tau(r, \sigma, \ell) \rc \label{locmomlum}
\eeq
where `${\rm cz}$' refers to the base of the convection zone.
$\tau$ corresponds to the integration of the local damping rate, and takes into account
the mean molecular weight stratification
\beq
\tau(r, \sigma, \ell) = \lc \ell(\ell+1) \rc ^{3\over2} \int_r^{r_c}
\lp K_T + \nu_v \rp \; {N {N_T}^2 \over
\sigma^4}  \left({N^2 \over N^2 - \sigma^2}\right)^{1 \over 2} {\diff r
\over r^3} \label{optdepth}
\eeq
(Zahn et al. 1997).
In this expression, ${N_T}^2$ is the thermal part of the Brunt-V\"ais\"al\"a frequency,
$K_T$ is the thermal diffusivity and $\nu_v$ the (vertical) turbulent viscosity.
$\sigma$ is the local, Doppler-shifted frequency
\beq
\sigma(r) = \omega - m
\lc \Omega(r)-\Omega _{\rm cz} \rc \label{sigma}
\eeq
and $\omega$ is the wave frequency in the reference frame of the convection
zone.
Let us mention that, in this expression for damping, only the radial velocity
gradients are taken into account. This is because angular momentum transport
is dominated by the low frequency waves ($\sigma \ll N$), which implies
that horizontal gradients are much smaller than vertical ones (\cf\,Eq.~\ref{kradial}).

When meridional circulation, turbulence, and waves are all taken into account,
the evolution of angular momentum follows
\beq
\rho \dtt \lc r^2 {\Omega}\rc =
\frac{1}{5 r^2} \drr \lc \rho r^4 \Omega U \rc
+ \frac{1}{ r^2} \drr \lc \rho \nu_v r^4 \dr{\Omega} \rc
 - \frac{3}{8\pi} \frac{1}{r^2} \drr{{\cal L}_J(r)},
\label{ev_omega}
\eeq
(Talon \& Zahn~1998)
where $U$ is the radial meridional circulation velocity.
This equation takes into account the advective nature of meridional circulation rather 
than modeling it as a diffusive process and assumes a ``shellular'' rotation (see 
Zahn~1992 for details).
Horizontal averaging was performed, and meridional circulation is considered only 
at first order. When we calculate the fast SLO's dynamics, $U$ is neglected in 
this equation. This is justified by the fact that, when shears are large such as in 
the SLO angular momentum redistribution is dominated by the (turbulent) diffusivity 
rather than by meridional circulation.
However the complete equation is used when secular time-scales are involved as required
when we compute full evolution models as in Charbonnel \& Talon~(2005).

\subsection{Shear layer oscillation (SLO) and filtered angular momentum luminosity}
One key feature when looking at the wave-mean flow interaction is that the dissipation of
IGWs produces an increase in the local differential rotation: this is caused by the
increased dissipation of waves that travel in the direction of the shear (see
Eqs.~\ref{optdepth} and \ref{sigma}). In conjunction with
viscosity, this leads to the formation of an oscillating doubled-peak shear layer that
oscillates on a short time-scale (Gough \& McIntyre~1998; Ringot~1998; Kumar,
Talon \& Zahn~1999). This oscillation is similar to the Earth quasi-biennial oscillation 
that is also caused by the differential damping of internal waves in a shear region.

This SLO occurs if the deposition of angular momentum by IGWs is large enough when 
compared with (turbulent) viscosity (Kim \& MacGregor~2001)\footnote{If viscosity 
is large, a stationary state can be reached.}.
To calculate the turbulence associated with this oscillation, we rely on
a standard prescription for shear turbulence away from regions with mean molecular weight gradients
\beq
\nu_v = \frac{8}{5} Ri_{\rm crit} K \frac{\lp r \diff \Omega/\diff r \rp ^2}
{{N_T}^2}
\label{maeder95}
\eeq
which take radiative losses into account (Townsend~1958;
Maeder~1995). This coefficient is time-averaged over a complete
oscillation cycle (for details, see TC05).

In the presence of differential rotation, the dissipation of prograde and retrograde 
waves in the SLO is not symmetric, and this leads to a finite amount of angular 
momentum being deposited in the interior beyond the SLO.
This is the filtered angular momentum luminosity ${{\cal L}_J}^{\rm fil}$.
Let us mention that in fact, the existence of a SLO is not even required
to obtain this differential damping between prograde and retrograde waves, and thus,
as long as differential rotation exists at the base of the convection zone, waves will
have a net impact of the rotation rate of the interior.

The SLO's dynamics is studied by solving Eq.~(\ref{ev_omega}) with small time-steps
and using the whole wave spectrum
while for the secular evolution of the star, one has to use instead the filtered 
angular momentum luminosity. Let us stress that in the case of the secular 
evolution we do not follow the SLO dynamics, because of its very short time scale.
Rather, we only consider the net angular momentum luminosity beyond the SLO, and 
its effect on chemicals is given by a local turbulence calculated from a study of 
the SLO's dynamic over very short time-scales.
Let us also mention here that, for both the SLO and the
filtered angular momentum luminosity, differential damping is required.
Since this relies on the Doppler shift of the frequency (see Eqs.~\ref{optdepth}
and \ref{sigma}), angular momentum redistribution will be dominated by
the low frequency waves that experience a larger Doppler shift, but that
is not so low that they will be immediately damped. Numerical
tests indicate that this occurs around $\omega \simeq 1~\mu {\rm Hz}$.

\section{The case of Pop~I low-mass stars}
A very important property of IGWs is that their generation and efficiency in
extracting angular momentum from stellar interiors depend on the 
structure of their convective envelope, which varies very strongly with the 
effective temperature of the star. 
Figure~\ref{VLiHyades} shows the $T_{\rm eff}$-dependence of the filtered angular momentum 
luminosity of waves below the SLO, which directly measures the efficiency of 
wave-induced angular momentum extraction, in zero-age main sequence stars around 
the Li dip. It appears that the net momentum luminosity slightly increases with increasing 
$T_{\rm eff}$, presents a plateau, and suddenly drops at the $T_{\rm eff}$ of the dip. 
This clearly indicates that the momentum transport by IGWs has the proper 
$T_{\rm eff}$-dependence to be the required process to explain the cool side of the 
Li dip (Talon \& Charbonnel 2003).

\begin{figure}
\centering
\includegraphics[width=4cm,angle=0]{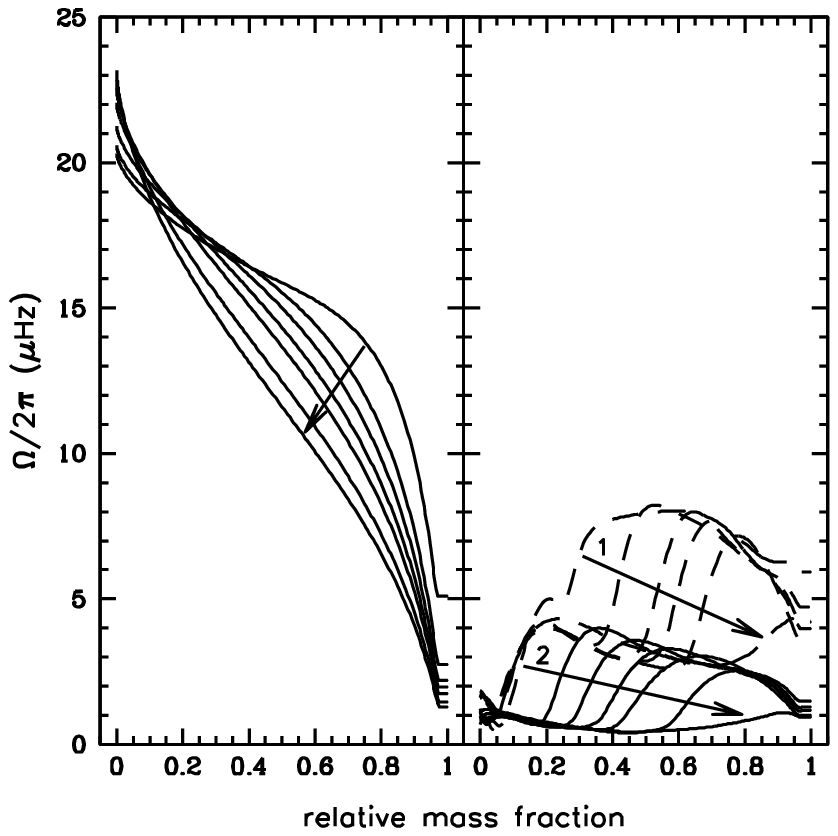}\qquad
\includegraphics[width=4cm,angle=0]{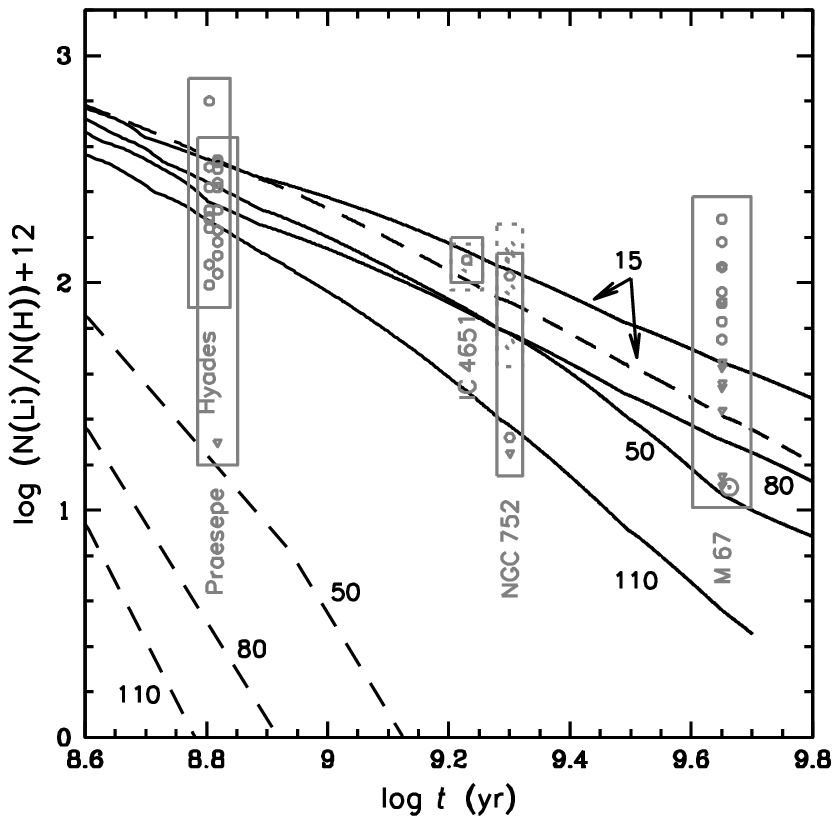}
  \caption{{\bf(Left)} Evolution of the rotation profile in a solar-mass model with and 
without IGWs.  The initial equatorial rotation velocity is $50~{\rm km\,s^{-1}}$, 
and identical surface magnetic braking is applied. ({\em left}) Model without IGWs. 
Curves correspond to ages of 0.2, 0.5, 0.7, 1.0, 1.5, 3.0 and 4.6 Gy that increase 
in the direction of the arrow. Differential rotation remains large at all times.
({\em right}) When IGWs are included, low-degree waves penetrate all the way 
to the core and deposit their negative angular momentum in the whole radiative region. 
Because the core's angular momentum is minute, it is spun down very efficiently. 
In the so-created ``slow'' region, damping of retrograde waves increases,
leading to the formation of a front, which propagates from the core to the surface.
Curves showing propagation of the first front (labeled 1) correspond to ages 
of 0.2, 0.21, 0.22, 0.23, 0.25 and 0.27 Gy. Further spin down leads to the formation 
of a second front (ages 0.5, 0.7, 1.0, 1.5, 3.0 and 4.6 Gy). The first front 
propagates faster than the second one due to stronger braking early in evolution.
At the age of the Sun, the radiative region is rotating almost uniformly. From 
Charbonnel \& Talon~(2005). 
{\bf (Right)} Evolution of surface lithium abundance 
with time for solar-mass stars. The vertical extent of boxes shows the range of 
lithium values as observed in various galactic clusters for stars with an effective 
temperature corresponding to that of the model $\pm100\,{\rm K}$ at the cluster age, plus 
a typical error in abundance determination. The horizontal extent corresponds
to the age uncertainty. Circles indicate abundance determinations, and triangles denote 
upper limits for individual stars. The solar value is shown with the usual 
symbol $\odot$. Solid lines correspond to models including IGWs and dashed lines to 
models without IGWs. Initial velocities are shown on the figure (in ${\rm km\,s}^{-1}$). 
In the cases without IGWs, except for the slowest rotator, lithium depletion is too 
strong, by orders of magnitude, at all ages. When included, IGWs, by changing
the shape of the internal velocity gradients, lead to a decrease of the associated 
transport of chemicals.
Lithium is then much less depleted and predictions account very well for the data. 
At all considered ages, the observed dispersion in atmospheric lithium is well 
explained in terms of a spread in initial velocities. 
From Charbonnel \& Talon~(2005). \label{evo_li}}
\end{figure}

Talon et al.~(2002) have shown, in a static model, that waves can efficiently 
extract angular momentum from a star that has a surface convection zone rotating 
slower than the interior.
Charbonnel \& Talon (2005) then calculated the evolution of the internal rotation profile 
for a solar-mass star with surface spin-down. We showed that, in that case, waves tend 
to slow down the core, creating ``slow'' fronts that propagate from the core to the 
surface (Fig.~\ref{evo_li}).  
These calculations confirmed, for the first time in a complete evolution of
solar-mass models evolved from the pre-main sequence to 4.6\,Gy, that
IGWs play a major role in braking the solar core (Charbonnel \& Talon~2005).
This momentum transport reduces rotational mixing in low-mass stars, leading to a 
theoretical surface lithium abundance in agreement with observations made in solar 
mass stars in open clusters of various ages (Fig.~\ref{evo_li}).

Figure~\ref{VLiHyades} shows our predictions for rotation velocities and
Li surface abundances together with the observed data at the age of the
Hyades. On the left side of the dip, IGWs play no role and
the predictions are taken from Charbonnel \& Talon (1999).
On the cool side of the dip IGWs are at act and lead to the rise
of the surface Li. The model at $\sim 5800~{\rm K}$ corresponds to a $1.0~M_{\odot}$
star. It was computed for 3 initial rotation velocities of $50$, $80$ and
$110~{\rm km\,s^{-1}}$ both in the case with (black dots) and without IGWs
(crosses). Models with IGWs are in perfect agreement with the observations,
both regarding the amplitude of the Li depletion and the dispersion
at a given effective temperature.
More details on these models will be given in Charbonnel et al. (in preparation).

\section{The case of Pop~II low-mass stars}
In the context of primordial nucleosynthesis, it has long been debated whether 
Pop~II stars could have depleted their surface Li abundance, just as their 
metal-rich counterpart did. 
Recent results on cosmic microwave background anisotropies, and especially those
of the WMAP experiment, have firmly established that the primordial Li abundance 
is $\sim$ 2.5 to 3 times higher than the measured Li value
in dwarf stars along the so-called Spite plateau (Charbonnel \& Primas 2005).
The main theoretical difficulty to reproduce 
these data is that the Li abundance
is remarkably constant in halo dwarfs, while it seems at first sight that depletion
would lead to a larger dispersion. 

A re-examination of Li data in halo stars available in the literature 
(Charbonnel \& Primas 2005) has led for the first time to a very surprising result:
the mean Li value as well as 
its dispersion appear to be lower for the dwarfs than for the subgiant stars.
In addition, all the deviant stars, i.e., the stars with strong Li deficiency and 
those with abnormally high Li content, lie on or originate from the hot side of the 
Li plateau. 
These results indicate that halo stars that have now just passed the turnoff have 
experienced a Li history sightly different from that of their less massive counterparts.

We suggested that such a behaviour is the signature of a transport process for 
angular momentum whose efficiency changes on the extreme blue edge of the plateau.
Such behaviour corresponds to that of the generation and filtering of IGWs in 
Pop~II stars (Talon \& Charbonnel 2004),
just as it does in the case of Pop~I stars. Indeed and as discussed previously, the 
generation of IGWs and, consequently, their efficiency in transporting angular momentum, 
depend on the structure of the stellar convective envelope, which 
in turn depends on the effective temperature of the star (Fig.~\ref{VLiHyades}). 
As in the case of Pop~I stars on the red side of the Li dip, the net angular luminosity 
of IGWs is very high and constant in Pop~II stars along the plateau up to 
$T_{\rm eff}\sim6300\,{\rm K}$. There, IGWs should dominate the transport 
of angular momentum and enforce quasi solid-body rotation of the 
stellar interior on very short timescale. As a result, the surface Li depletion is 
expected to be independent of the initial angular momentum distribution, implying 
a very low dispersion of the Li abundance from star to star.
In more massive stars however the efficiency of IGWs decreases and internal 
differential rotation is expected to be maintained under the effect of meridional
circulation and turbulence. Consistently, variations of the initial angular momentum 
from star to star would lead to more Li dispersion and to more frequent abnormalities 
in the case of the most massive stars where IGWs are not fully efficient, as required 
by the observations. We note that the mass-dependence of the IGWs efficiency 
also leads to a natural explanation of fast horizontal branch rotators. 
The proper treatment of the effects of IGWs together with those of atomic diffusion, 
meridional circulation and shear turbulence has now to be undertaken in fully consistent
models of halo stars. 

\section{IGWs in intermediate-mass stars}
Although waves produced by the surface convection zone can be ignored safely for 
more massive stars (i.e., with $T_{\rm eff} \geq 6700~$K) while on the main sequence, 
it is not the case for later evolutionary stages. In particular, Talon \& Charbonnel 
(2008) showed that angular momentum transport by IGWs emited by the convective envelope 
could be quite important in intermediate-mass stars on the pre-main sequence, 
at the end of the sub-giant branch, and during the early-AGB phase. 
This implies that possible differential rotation, which could be a relic of the 
star's main sequence history and subsequent contraction, could be strongly reduced 
when the star reaches the AGB-phase. This could have profound impact on the 
subsequent evolution. In particular, this could help solving the long-standing 
problem of the production of the s-process elements (Herwig et al.~2003; 
Siess et al.~2004), as well as explaining the observed white dwarf spins 
(Suijs et al.~2008). Work is in progress to quantify these effects.

Let us also mention that waves can be excited at the boundary of the convective core. These
waves travel towards the surface, and are fully damped there, as thermal diffusivity is quite large at
the star's surface, and could have a large impact on rotation profiles there (Pantillon, Talon, \& Charbonnel 2007). 
As these stars are fast rotators, the Coriolis force must be accounted for, and much work remains
to be done in that direction (Pantillon et al. 2007; Mathis et al. 2008).

\section{Conclusions}\label{sec:concl}

IGWs have a large impact on the evolution of low-mass stars, especially through 
their effect on the rotation profile, which then modifies meridional circulation 
and shear turbulence. 
Within this framework, the hydrodynamical models including the combined effects of
meridional circulation, shear turbulence and internal gravity waves (using an 
excitation model that reproduces the solar p-modes) successfully shape
both the rotation profile and the time evolution of the surface lithium abundance
in low-mass stars. Up to now, no other theoretical model has achieved such a goal.
Our comprehensive picture should have implications for other difficult unsolved
problems related to the transport of chemicals and angular momentum in stars. 
We think in particular to the stars on the horizontal and asymptotic giant branches 
that exhibit unexplained abundance anomalies.
No doubt that all these so-called ``non-standard" physical processes must 
be part of the art of modelling stars in the 21st century.

\begin{discussion}

\discuss{Woitke}{Why do the simulations of Glatzmaier et al. for the excitation of 
gravity waves by convection fail? Do they yield too strong or too weak convection?}

\discuss{Charbonnel}{The convective motions in this simulation are ``too lazy". 
The numerical resolution is too small to resolve the hammering of plumes properly.
Prandtl-numbers are too low by orders of magnitudes.}

\discuss{Kupka}{I would like to comment on the question of the resolution and 
turbulence in 3--D global solar simulations. If you take the case of $512^3$ grid points, 
at the bottom of the solar convection zone you have a horizontal resolution of 
$\sim$6000~km. The local pressure scale height there is $\sim$50000~km. So if you 
consider the energy carrying scales to be of that size, horizontally 
$R_{eff} \sim (L/R)^{4/3} \sim 20$. Vertically the resolution is perhaps some 600~km, 
hence $R_{eff} \sim 370$. Also, for comparison, the so-called ``extent of 
overshooting", according to helioseismology, is some 2500~km, if you take it 
to be $\sim 0.05H_p$. Thus, the simulations cannot resolve shear-driven turbulence
created by the flow on such scales. It is numerically simply too expensive to do that
on computers currently available.}

\discuss{Langer} {You prefer g-modes as mechanism to slow down the core of MS stars, 
since this gives Li-depletion on the cool side of the dip but not on the hot side.
Would you not expect a similarly different effect of internal magnetic transport 
for both sides of the dip, since the cool stars suffer magnetic braking but the
hot stars don't?}

\discuss{Charbonnel}{The question is more related to the transport of angular 
momentum inside the star, and not to the braking at the surface. For the moment 
there is no argument to suspect a $T_{\rm eff}$-dependence of the angular momentum transport
by magnetic field.}

\discuss{Ludwig}{The Li abundance on the Spite plateau is very homogeneous over 
almost 1000~K in $T_{\rm eff}$ and 2 orders of magnitude in metallicity. The production 
rate of gravity waves, and I presume its mixing is dependent on the mass of the
convective envelope. Can you comment how the homogeneity of the Li abundance is
consistent with the substantial dependence of the stellar structure.}

\discuss{Charbonnel}{There is in fact a threshold value for the momentum luminosity
of waves above which the waves are extremely efficient in transporting angular 
momentum. The dwarf stars that lie on the Li plateau, despite their slightly 
different internal structure, all managed to be above that threshold in their 
infancy. We thus expect that they all managed to become solid-body rotators.
As a consequence, they must have depleted Li in a very homogeneous way.}

\end{discussion}

\end{document}